\documentclass[nohyper]{JHEP} 

\usepackage{epsfig}

\def\desepsf(#1 width #2){\epsfxsize=#2 \epsfbox{#1}}
\def \pom {{\scriptscriptstyle \kern -0.1em I \kern -0.25em P}}

\title{\boldmath $Q {\overline Q} g$ contribution to diffractive 
$J/\psi$ electroproduction}

\author{F. Hautmann\\
	Department of Physics, Pennsylvania State University\\  
        University Park PA 16802}

\abstract{
We  study  
 the diffractive electroproduction 
of quarkonia   from  
  quark-antiquark-gluon states in  the photon  
wave function.  We show that these   
states contribute  to the  
 leading-power and leading-logarithm level. 
We suggest the 
 measurement of 
 $J/\psi$ production via inelastic diffraction 
to study  color-transparency 
and color-opacity effects in the diffractive gluon 
distribution. }


\preprint{PSU-TH/239}

\begin{document} 

\section{Introduction}

$J/\psi$ mesons are produced copiously via  hard-diffractive reactions 
 in  lepton-hadron collisions at  high energies. 
In these processes, the initial 
hadron emerges intact in the final state and 
the meson   is produced via a short-distance interaction, 
involving momentum transfers of the order of the heavy quark mass 
or, for  highly off-shell photons, the photon virtuality.

The 
elastic   
process $ \gamma + A \to V + A^\prime$, where 
 $A$ and $A^\prime$ are the initial and 
final hadrons and  $V$ is the quarkonium, 
has been studied extensively at HERA~\cite{zeusexcl,h1chmn,h1excl}. 
This process is computable in QCD~\cite{rys,brodetal,fskoepf} 
in the sense that a short-distance coefficient, calculable 
as a power expansion in $\alpha_s$,  can be factorized~\cite{cfsfact} 
from  a parton density describing the long-distance dynamics  
of the hadronic  state $A$    
and a wave function describing 
the long-distance dynamics  of the produced bound  state. 
A typical leading-order graph is shown in Fig.~1. 
The parton density probed by this process is a nondiagonal 
parton  density~\cite{cfsfact,leip,xdji,radyu}:  
\begin{equation}
\label{glueskewed} 
\phi_{g} (\xi , x_\pom, t, \mu ) = {1 \over 2\pi  
\xi p_{\!A}^+} \int d y^-
e^{i \xi p_{\!A}^+ y^-} 
\langle A^\prime |\widetilde G_a(0)^{+j}
\widetilde G_a(0,y^-,{\bf 0})^{+j}| A \rangle 
\hspace*{0.4 cm}  ,    
\end{equation}
where  $p_A^+$ is the lightcone plus momentum of hadron $A$, 
 $x_\pom=1-p_{A^\prime}^+/p_A^+$,   
  $t = (p_A - p_{A^\prime})^2$.  The 
operator   
$\widetilde G (0,y^-,{\bf 0})$ is the gluon field strength multiplied 
by a path-ordered exponential of the line integral of 
the color potential  from $y^-$ to $\infty$. The ultraviolet divergences 
from the operator product are 
   renormalized at the scale~$\mu$.

In this paper we are  concerned with  the production of quarkonia by 
   inelastic diffraction,  
$ \gamma + A \to V + X + A^\prime$, a process that   has not 
been considered  so far but could be 
measured at HERA and at future 
lepton-proton and lepton-nucleus  colliders. 
This process probes a different parton density than (\ref{glueskewed}): 
see  Eq.~(\ref{opg}) ahead. 
As we will see,  for sufficiently 
 large values of the  hard-scattering scale  
 its cross section 
 is enhanced by two powers of the hard scale   
compared to the case of Fig.~1.

In the rest frame of the target (see Sec.~2), 
the elastic production  is dominated by 
fluctuations of the incoming photon 
into  quark-antiquark  states, while 
the process we consider in this paper   is dominated by 
fluctuations 
into  quark-antiquark-gluon  states.  
The main interest in measuring this process is to observe 
effects of the coupling of these  states to  the 
target's soft color field.  The color dipole interaction  
  is  stronger  for 
  $Q {\overline Q} g$ states than for 
  $Q {\overline Q}$ states. 
 We   discuss  consequences of this    in Sec.~5. 

The contents of the paper is as follows.  
We present the physical picture of the 
process in Sec.~2. This discussion uses the target  rest frame 
 and gives  
an ``$s$-channel view'' of the process, based on the 
evolution 
of the system into which the incoming photon dissociates. 
 Then we relate the evolution  of this system  to 
diffractive matrix elements of gluon field operators,  
and in Secs.~3 and 4 we  give 
 factorization formulas  for  the 
quarkonium  
  cross sections. This formulation corresponds to     a 
  ``$t$-channel view'' of the process. 
  We use these results to contrast   the   scaling behaviors 
  of the elastic  and inelastic  contributions.   
   Sec.~5 contains a  concluding 
 discussion, focusing on  
 the possibility that 
 diffractive interactions of $Q {\overline Q} g$ 
 states with the target be dominated by small transparency lengths, and 
  on the implications of this for  the diffractive gluon distribution.  
The analysis of this paper applies to both $J/ \psi$ and $\Upsilon$ 
mesons.  
Numerical evaluations of the cross sections are left to 
future work.

\section{\boldmath  
$J/\psi$ production by $Q {\overline Q} g$ states: the $s$-channel 
point of view}

In this section we work in 
the rest frame of the target  and 
present the physical picture for  
the production of quarkonia  from 
quark-antiquark-gluon states in the photon  wave function. 

As is well known, the spacetime structure~\cite{bjrest,niko} 
of hard-diffractive  processes in 
electroproduction 
 looks especially simple 
in the target rest frame. 
In this frame, at  
lightcone times far in the 
past a highly energetic  photon 
dissociates into a partonic system. 
 Much later,  
  this  system interacts with the color field 
of the target, mainly through its color dipole  
moment~\cite{niko}.  
The 
physics of the parton densities is re-expressed in this frame 
in terms of 
 the dipole-target scattering 
  (see reviews in~\cite{muedis98,ffss99}). 
The elastic production of quarkonia (Fig.~1)  has been  analyzed 
in this context  in Refs.~\cite{rys,brodetal,fskoepf}. 
This process  is dominated by  the contribution of the 
quark-antiquark component  in the 
photon wave function. 

Fig.~2 depicts a 
 typical  inelastic 
contribution to the diffractive production of quarkonia. 
  In the target rest  frame,  
a quark-antiquark-gluon state is created by the photon 
well outside of the target hadron. 
This system has large minus momentum
 and  travels a long 
distance before  interacting with the target's color field. 
(This interaction is 
schematically represented in Fig.~2 
as  two-gluon exchange.  In general, it  
includes any number of soft gluons and is nonperturbative.  
 Note also that  contributions in which $k_1$ and $k_2$ 
attach to the quark lines in the upper subgraph 
are     
 to be 
considered together with the contribution depicted in Fig.~2.   
One needs to include them  
to construct   gauge-invariant quantities.)  
Over long times, the quark and antiquark bind into a 
quarkonium. 
The emission of two gluons  into the 
final state in Fig.~2 is required by color conservation. 
The interaction of the fast-moving parton  system with the 
target's field  leaves  hadron $A$ almost intact, and      
 produces   a final state consisting of the quarkonium plus jets  
plus the diffracted hadron. 

\FIGURE{\desepsf(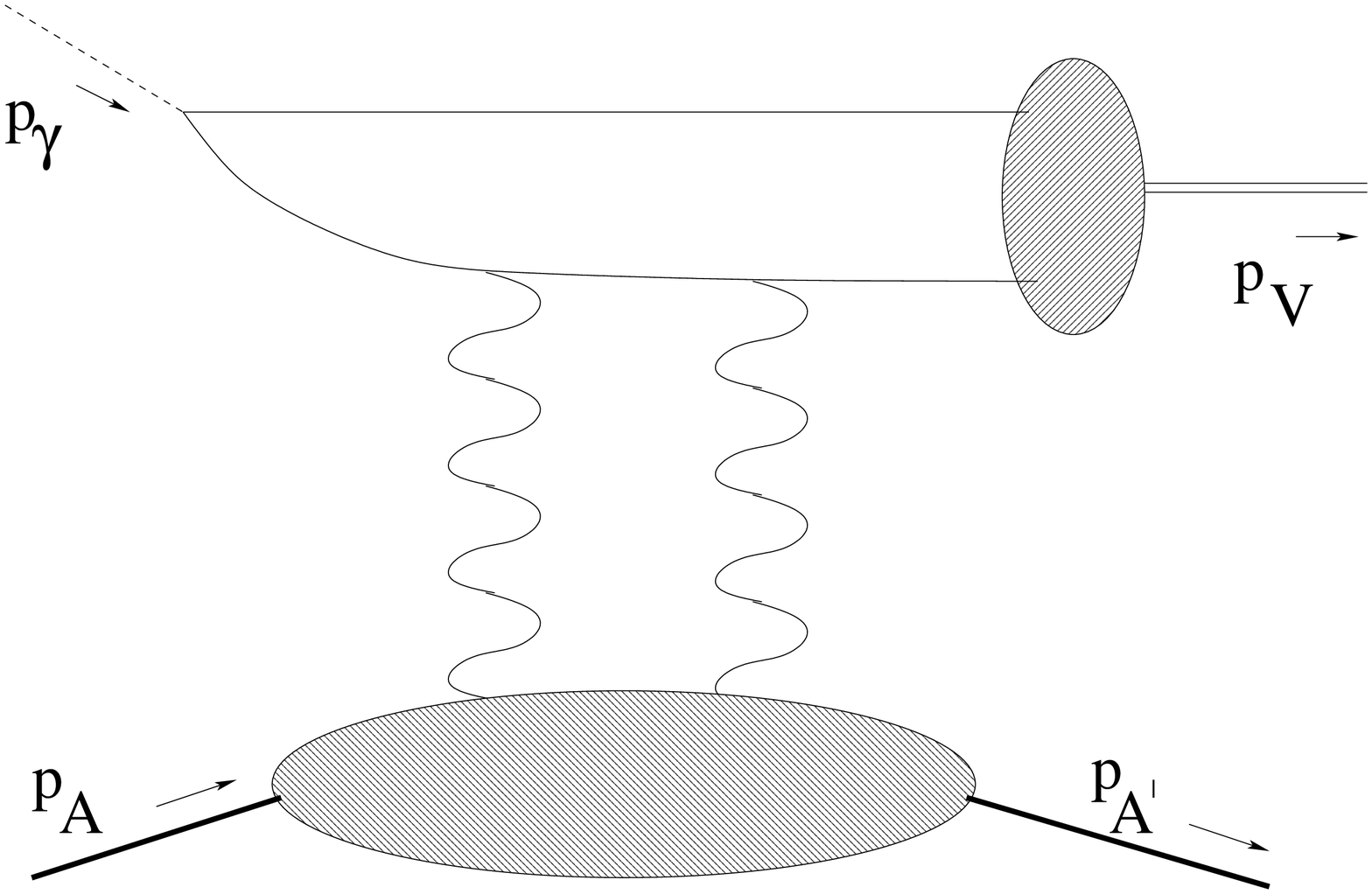 width 9 cm)
\caption{A typical graph  for the elastic process 
$ \gamma + A \to V + A^\prime$. }
\label{firstfig}
}

\FIGURE{\desepsf(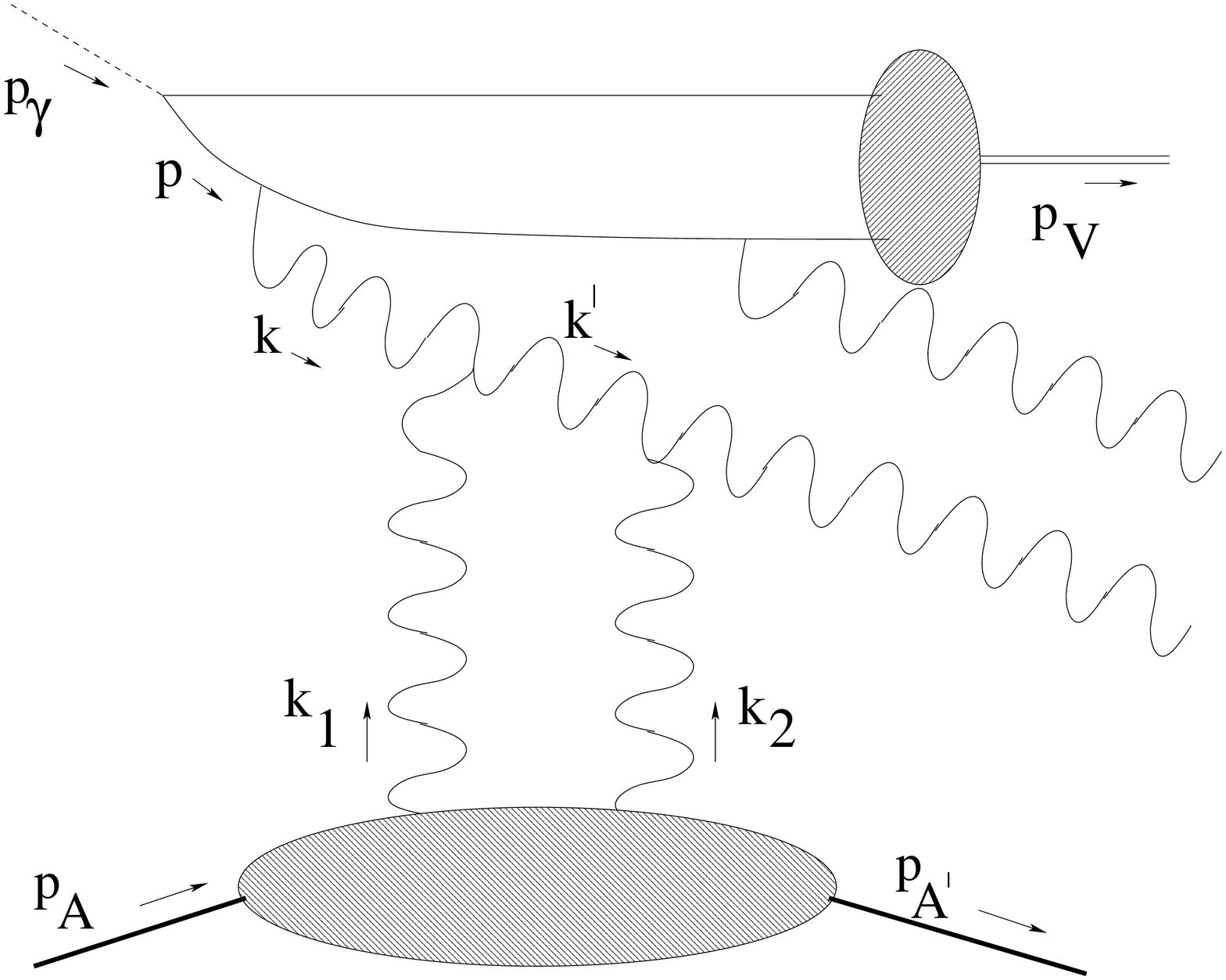 width 9 cm)
\caption{$Q {\overline Q} g$  contribution to diffractive 
quarkonium  production. }
\label{secondfig}
}

In the long interval between the formation of the parton 
system and the interaction with the target's  field,  gluons can be 
radiated, possibly at small lightcone separations. These graphs  
contribute to  evolution 
and radiative corrections.  
   We will see later 
 the result of  including them.

Consider  now 
the upper subgraph in Fig.~2, in which $k_1$ and 
$k_2$ couple  to the $Q {\overline Q} g$ system.
This subgraph 
is dominated by 
 the 
 region in  which 
 the transverse momentum $ | {\bf p} |$ is of order $M$ 
 (with $M$ the 
heavy meson mass) and much larger than the gluon transverse 
momenta $ |{\bf k}|$ and $|{\bf k}^\prime |$.  
This region contributes a hard-scattering 
squared amplitude of  order $\alpha \alpha_s^2$, that is,   
   of the same order  
 as the hard scattering  in 
Fig.~1.  By power counting in the coupling,   
the  contributions in Figs.~1 and 2 are of the same size. 
Later on we will see  that 
the contribution in Fig.~2 dominates the contribution in Fig.~1 
by power counting in the hard-scattering scale.

The observation above indicates that in the dominant 
region  the gluon $k$ goes 
near the mass shell.  The process becomes sensitive to 
 long distances, and involves the 
nonperturbative  distribution of gluons in the hadron $A$, 
subject to   
 the condition that the hadron is diffractively scattered. 
This distribution is  depicted in 
Fig.~3. It is defined in terms of the 
   gluon operator ${\widetilde 
G}$ introduced in  Eq.~(\ref{glueskewed})  
by the matrix elements  
\begin{eqnarray}
\label{opg}
&& {{d\, f_{g} } \over
{dx_\pom\,dt}} (\beta , x_\pom , t , \mu ) =  
{1 \over (4\pi)^3 
\beta x_\pom p_{\!A}^+}\sum_{X} \int d y^-
e^{i\beta x_\pom p_{\!A}^+ y^-} 
\nonumber\\
&&   \times 
\langle A |\widetilde G_a(0)^{+j}
| A^\prime, X \rangle 
\langle A^\prime, X|
\widetilde G_a(0,y^-,{\bf 0})^{+j}| A \rangle 
\hspace*{0.2 cm} ,     
\end{eqnarray}
with $\beta x_\pom$  the fraction of the hadron's plus momentum 
 carried by the gluon. 
These matrix elements have been studied 
in detail in \cite{hksnpb,bere}. 
The vertex at the top of Fig.~3 represents 
the gluon operator in Eq.~(\ref{opg}), including a color-octet 
eikonal line defined by the path-ordered 
exponential in 
$\widetilde G$~\cite{hksnpb,bere}. 
The eikonal line can be interpreted physically as a 
stand-in for the quark-antiquark pair that was produced 
by the electromagnetic current but is too compact to  be 
resolved by the hadron's color field. 
Attachments of gluons to this line 
are incorporated as explained in  \cite{hksnpb}.

\FIGURE{\desepsf(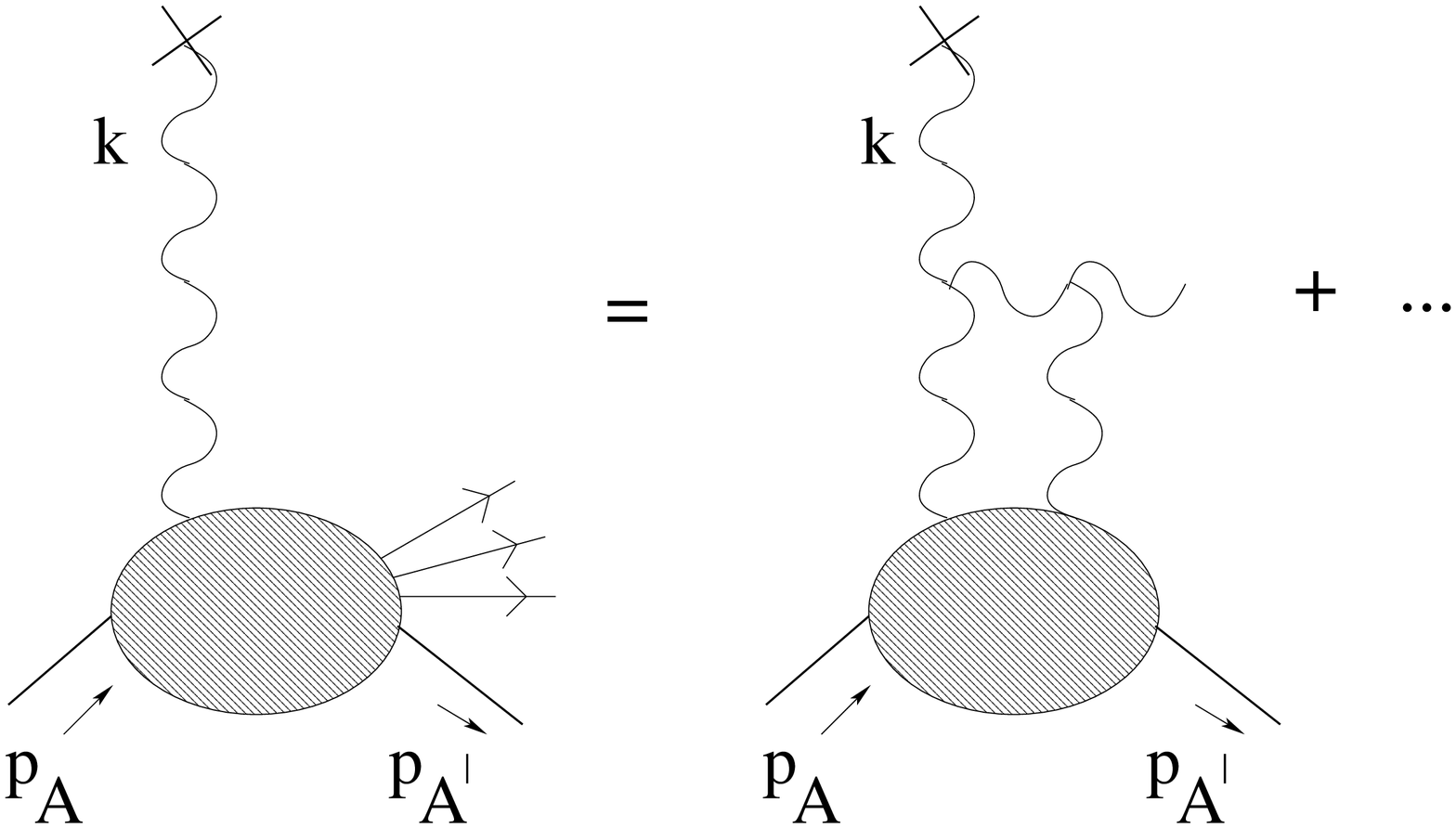 width 9 cm)
\caption{Diffractive matrix element for the gluon operator 
(Eq.~(\ref{opg})). The graph  
 on the right hand side corresponds 
to the contribution  depicted in Fig.~2.}
}

\section{\boldmath  
 The $t$-channel 
point of view}

The precise relation of the vector meson  
cross section with the matrix elements (\ref{opg}) 
is given by  a factorization formula. This 
can be obtained by applying to the graph of Fig.~2 
standard  methods~\cite{libby,csrev}  for   
 determining  the  leading-power asymptotics  in the limit 
of a large hard-scattering scale.

The 
 leading integration regions in momentum space    
  involve (see Fig.4)   
i) a 
beam jet  in the 
direction of hadron $A$,  including 
the final hadron $A^\prime$; ii) a hard subgraph, 
in which 
all virtual lines are off shell by order $M$; 
iii) final states jets, including the vector meson, that are not 
in the direction of $A$; iv) a soft  subgraph, consisting of 
soft gluons  that link the beam jet with the hard 
subgraph and final state jets.

The method  \cite{libby,csrev} instructs us to 
a)  use Ward identities to 
factor 
out 
soft gluon subgraphs  from final-state  and beam 
jets, and  b)  
   sum 
 over real and 
virtual graphs  to show that all soft gluon exchanges  
(dashed lines in Fig.~4) 
cancel in the cross section. 
This can be accomplished in a standard manner   
in the problem at hand, 
except for the treatment of soft gluons coupling to the beam 
jet in the plus  direction:   
owing to  the diffractive requirement 
on the final state, 
there are both initial-state and final-state couplings, giving  
  poles  
 on both sides of the integration contour in the 
complex  plane of the  gluon minus momentum $r^-$. This  
 prevents us from 
deforming the $r^-$ contour to the region where the soft 
approximation is valid.  

\FIGURE{\desepsf(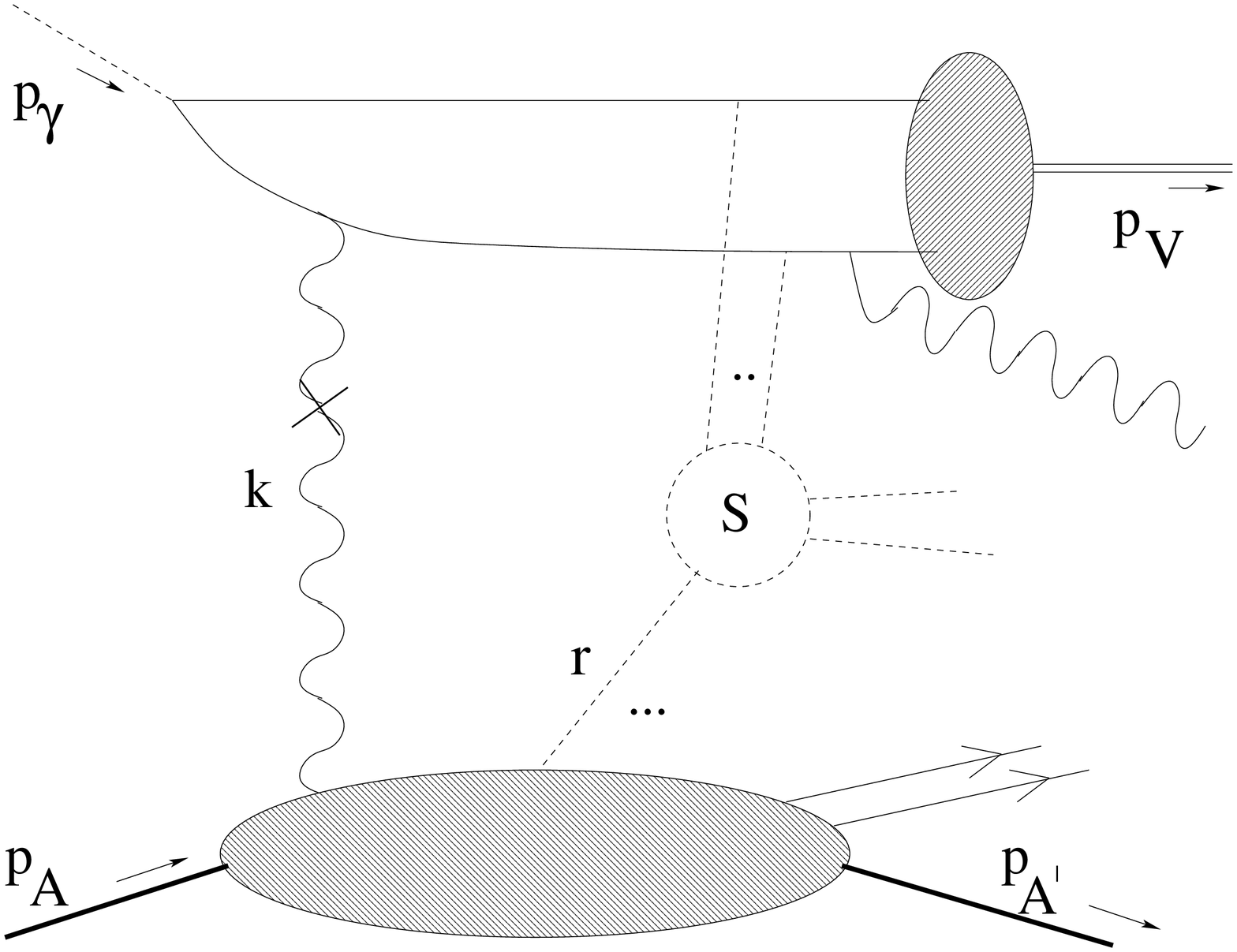 width 9 cm)
\caption{Factorization 
of the inelastic diffraction process. 
Soft gluon exchanges (dashed lines) cancel.}
}

This obstruction, however, can be overcome in the 
same way as in  diffractive deep inelastic 
scattering~\cite{diffcontour}, by going to the  
complex plane of the  plus momentum 
component  $r^+$, and showing that in this plane 
the contour is not trapped by singularities in the 
beam jet subgraph. Then the cancellation of soft gluon 
exchanges goes through, and  
 factorization follows. 
 To keep the notation simpler we first 
consider 
the case  
$Q^2 = 0$, limiting 
ourselves to the direct-photon part of photoproduction. 
(We comment on the resolved-photon part below.) 
The 
factorization formula 
for the $ \gamma + A \to V + X + A^\prime$ cross section 
reads   
\begin{equation} 
\label{photofact} 
{ { d \sigma^{(photo)} 
 } \over { dx_\pom\, dt}} (M^2 , W^2 ,  x_\pom , t)  
= \sum_{a = q , {\bar q}, g} \, \int_{\omega/x_\pom}^1  
d \beta   \, { df_{a} 
\over dx_\pom\, dt} (\beta,x_\pom,t,\mu^2) \, 
 {\hat \sigma}_{ a} ( {\omega \over {\beta x_\pom}} , M^2 ,   \mu^2 ) 
\hspace*{0.3 cm} 
\end{equation}
with $W^2= (p_A+ p_\gamma)^2$, 
$ \omega = M^2 /  W^2$. The coefficient ${\hat \sigma}_{ a}$ 
describes  the meson  production subprocess represented by  the 
upper subgraph in Fig.~4. 
In the nonrelativistic expansion for the bound state, 
${\hat \sigma}_{ a}$  
factorizes into the product of a perturbative factor, 
describing the production of the quark-antiquark pair, and 
a nonperturbative factor, describing the binding of the 
quark and antiquark into a quarkonium.  
In this paper  we  consider  only 
the lowest order  
 in this expansion, corresponding to  
color-singlet quark-antiquark 
 states~\cite{psiphoto}.  
 We comment below on the possible role of 
 color-octet states.

Both ${\hat \sigma}_{ a}$ and $f_a$ in Eq.~(\ref{photofact}) depend on the 
factorization scale $\mu$.  As noted earlier, 
the evolution in $\mu$ arises, via renormalization of the 
ultraviolet divergences,  
from  gluons being 
radiated from the top subgraph in Fig.~2. 
We refer the reader to 
Refs.~\cite{hksnpb,bere,diffcontour,cfsfact}   
for detailed treatments  of evolution in hard-diffractive processes.

To the leading order in $\alpha_s$, only 
the term with $a = g$  enters in the sum in Eq.~(\ref{photofact}).  
This is the term depicted in Figs.~2,4. 
 Compare this with the case of the 
diffractive structure function $F_2$, where $a = g$ enters to the 
next-to-leading order, and of diffractive jet production, where 
both $a = g$ and $a = q $ enter to the leading order. 
(See, e.g., \cite{niko94,barniko} for early treatments of gluon radiation 
in these processes.) 
This makes the  quarkonium  cross section  an especially  
important 
experimental probe of the physics of  large gluon densities in 
 diffraction.

The  coefficient ${\hat \sigma}_g$ 
that relates the   quarkonium   cross section 
 to the diffractive gluon density 
can be obtained from the  
perturbative calculations of \cite{psiphoto} and  reads  
\begin{eqnarray} 
\label{sighat} 
 {\hat \sigma}_{g} ( \omega , M^2, \mu^2) &=&  
{32  \over { 3  \pi}} \ 
{{  \alpha \ e_Q^2 \ \alpha_s^2 (\mu^2)  } \over {  M^2} }
\ {{ | \psi (0) |^2  } \over {  M^3 } }  \ \omega 
\nonumber\\
&& \times 
 \left\{ 
{ { 1 + \omega} \over { 1 - \omega} } + 
{ { 1 - \omega} \over { (1 + \omega)^2} } 
+ 2 \omega \ln {1 \over \omega} \ \left[ 
{ 1 \over { (1 + \omega)^3} }  - { 1 \over { (1 - \omega)^2} } 
\right] \right\} 
\hspace*{0.3 cm} . 
\end{eqnarray}
Here $ \psi (0)$ is the quarkonium wave function at the origin.   
Eq.~(\ref{sighat})  
 vanishes like $\omega$  for large energies   
($\omega \to 0$) and  like $(1- \omega)$ near threshold 
($\omega \to 1$). In these boundary regions 
 higher order  corrections~\cite{zerw} to the  coefficient 
 become  particularly important. 
Notice the behavior of the cross section  (\ref{photofact}) 
in the hard scale $M^2$. 
Apart from the dimensionless factor $ |\psi (0)|^2 / M^3$,  coming  
from 
the nonrelativistic treatment of the bound state, 
$d \sigma / dt $ goes like 
 $  1 / M^2$.  This is to   be contrasted  with the 
behavior 
$ 1 / M^4$~\cite{rys,fskoepf} 
of the elastic process (Fig.~1).  That is,  for 
sufficiently large $M$   
the cross section is 
significantly larger than its elastic part.

As noted above, this treatment does not include 
terms from color-octet quark-antiquark 
 states.  Color-octet terms  can in general  be    
 large,   because, although 
suppressed by powers of the  heavy quark velocity, they are enhanced 
by a power of $1/\alpha_s$~\cite{caccia}. In this respect, they   are  more 
important in the inelastic-diffraction process than in the elastic case. 
Observe however that 
if the final state is diffractive 
the contribution of  color-octet states does not necessarily follow  
the simple factorized form that is 
commonly used in applications of the nonrelativistic expansion. 
The  soft gluons emitted in  
 the transition of the color-octet quark pair to quarkonium 
may couple to the diffracted final state. 
These soft gluons are likely to 
   reduce the probability  for producing  a 
quarkonium  via  a color-octet pair   while 
  leaving  hadron $A$   intact. 
The net effect is 
that   color-octet contributions  should be suppressed in diffractive 
production compared to expectations 
valid in the fully inclusive case.

The resolved-photon part of photoproduction cannot be described 
by a factorization formula of the kind (\ref{photofact}), because 
the 
possibility of color being exchanged with the 
 photon beam jet spoils the cancellation of the 
 soft gluon subgraphs.   
 Since the direct and resolved parts cannot be 
 separated in general beyond the leading logarithms, this limits 
 the applicability of factorization in the  $Q^2 = 0$ case.  

 Note however that, if we 
 estimated     
the sign and  order of magnitude of the soft contributions 
from diffractive  hadron-hadron  data (see, e.g.,  \cite{albrow}), this   
 would suggest that  the relative importance of the resolved part 
compared to the direct  part is smaller in  diffractive 
photoproduction than in inclusive photoproduction, perhaps by 
a  factor of several. 
This, combined with the use of kinematic cuts designed to suppress 
the resolved process, may allow one to make a meaningful test of the direct 
process. 

\section{\boldmath    $Q^2 \neq 0$ case } 

The leptoproduction case $Q^2 \neq 0$ is conceptually simpler than 
 photoproduction, because it does not involve the 
processes with soft color exchange discussed at the end of the 
previous section. In the leptoproduction case 
a result of the kind  (\ref{photofact}) holds. More precisely, 
both the longitudinal and the transverse cross sections 
satisfy a factorization formula in terms of the diffractive 
matrix elements (\ref{opg}):  
\begin{eqnarray}  
\label{leptosig} 
&& { { d \sigma^{(lepto)} 
 } \over { dx_\pom\, dt \ dQ^2 \ d y }} (M^2 , S ,  Q^2 , 
y , x_\pom , t) = 
{ \alpha \over { \pi  }}  { 1 \over {y Q^2}} 
\\
&& \times 
\left[ 
(1-y+{y^2 \over 2} ) { { d \sigma_2 
 } \over { dx_\pom\, dt \ }} (M^2 ,  Q^2 ,  y S  ,
 x_\pom , t) 
- {y^2 \over 2} { { d \sigma_L 
 } \over { dx_\pom\, dt \ }} (M^2 ,  Q^2 ,  y S  ,
 x_\pom , t) \right] 
\hspace*{0.1 cm}  , 
\nonumber\\
\nonumber      
\end{eqnarray} 
\begin{equation} 
\label{leptofact} 
{ { d \sigma_i 
} \over { dx_\pom\, dt}} 
= 
\sum_{a } \, \int   
d \beta   \, { df_{a} 
\over dx_\pom\, dt} (\beta,x_\pom,t,\mu^2) \, 
 {\hat \sigma}_{i, a} 
( {\omega^\prime \over {\beta x_\pom}} , M^2 ,   Q^2 , \mu^2 ) 
\hspace*{0.9 cm}  ( i = 2 , L ) \hspace*{0.2 cm} . 
\end{equation} 
Here, in standard notation, 
 $ \sigma_2$ is the sum of the transverse and longitudinal cross 
sections, $\sqrt{S}$ is the lepton-hadron center-of-mass energy, 
$y$ is the lepton energy loss in the target rest frame, 
and $\omega^\prime = (M^2 + Q^2) / (y S)$.  

Note that in the case of the elastic 
 process (Fig.~1), in contrast,   
 a factorization 
formula    exists  so far  only for the longitudinal 
polarization~\cite{cfsfact}. 

The coefficients 
 ${\hat \sigma}_{i, a}$  generalize  
   Eq.~(\ref{sighat}) to $Q^2 \neq 0$. 
Consider the power counting at large $Q$. 
(By comparison,  recall~\cite{brodetal,fskoepf}  
that the elastic process of Fig.~1 scales like  $1/Q^6$ at large $Q$.)   
 For $Q \gg M$, 
a factor 
of $1 / Q^2$ 
replaces the $1 / M^2$ of  Eq.~(\ref{sighat}). 
Fig.~2 has a suppression factor $M^2 / Q^2$  
for producing a quarkonium of mass $M$ when the 
momentum flowing in the  quark-antiquark subgraph 
is of order $Q$. 
However,  note that for  
$Q \gg M$ Fig.~2  is 
no longer  dominant,  as 
contributions from the intrinsic heavy-quark distribution and from 
parton fragmentation into quarkonium 
scale like $1  / Q^2$. 
We therefore  suggest considering 
the region of $Q$ of 
order $M$,  or smaller than $M$. 
In this region,  quark-antiquark-gluon states give a contribution to 
diffractive quarkonium production that is both leading 
in powers of the hard scale and  leading in the running coupling. 

 Detailed  studies are warranted  
to   assess the numerical importance  of  this contribution 
at HERA and future $e A$ colliders. 

\section{Discussion }

 We have seen that the  
electroproduction of quarkonia  by inelastic diffraction  
measures, through the factorization formulas of Secs.3 and 4, 
the  diffractive gluon distribution (\ref{opg}). 
For $Q$ of order $M$,  or smaller than $M$,   
 this process provides a remarkable   probe of   
  quark-antiquark-gluon 
 states in the photon lightcone wave function. 
Recall that, in contrast, 
the production of quarkonia by   elastic diffraction 
couples predominantly to quark-antiquark states.

The differences between the dynamics   of $Q {\overline Q}$ and 
$Q {\overline Q} g$ 
 states   
 scattering   off the hadron's soft color  field  
 have  been investigated 
extensively in the 
color dipole approximation (see, e.g., 
\cite{niko,muedis98,ffss99},  and 
references therein).  
Color-octet dipoles interact  more strongly with the target. As a 
result,   
 the probability for the parton system  to get through the 
hadron without breaking it up dies off faster  as the dipole size 
increases, and  the hadron starts to look opaque earlier.

Based on the analysis of  these   
 effects,   a  physical picture 
 for the diffractive parton distributions 
was suggested in~\cite{hscoltrans}.   
 One of the implications of  the onset of 
  color opacity  is that,  
 if the maximum parton separation $1/\kappa$ for which the hadron is 
transparent is sufficiently small, 
a perturbative calculation~\cite{hkslett}  of the 
  $\beta$ dependence of the diffractive  distributions 
should apply.   Such a calculation is incorporated in~\cite{hscoltrans} 
and in different 
models~\cite{buchmue,golec} for the diffractive distributions. 
The overall normalization of the distributions 
is   inversely  proportional to the 
square  of the transparency length $1/\kappa$ 
(in units of the hadron radius $r_A$), while their  shape in $\beta$ 
is  given by perturbatively computable functions $\varphi$: 
\begin{equation}
\label{gform}
f_{g}   \propto (N_c^2 -1) \ \kappa_g^2  r_A^2 \ 
 \varphi_g (\beta) 
\hspace*{0.1 cm}  , 
\hspace*{0.5 cm}
f_{q}   \propto N_c \ \kappa_q^2  r_A^2 \ 
 \varphi_q (\beta) 
\hspace*{0.1 cm}  .        
\end{equation}

In~\cite{hscoltrans}    predictions based on  these distributions 
were compared 
with the HERA data~\cite{zeusf2}  
for  the  diffractive structure function $F_2$. 
The comparison 
  indicates that 
 transparency lengths are 
  smaller for $Q {\overline Q} g$ states 
than  for  $Q {\overline Q}$ states. 
The  $F_2$ data are consistent with 
    $\kappa_g / \kappa_q \approx C_A / C_F$. 
 Note that, together with the color factors explicitly shown in 
(\ref{gform}), this value gives back 
 the result~\cite{hkslett} 
for  the  ratio of the diffractive 
 gluon and quark distributions in a small dipole,  
\begin{equation}
\label{colratio} 
{f_g \over f_q} \propto {{ C_A^2 (N_c^2 -1) } \over  { C_F^2 N_c }} 
= {27 \over 2} 
\hspace*{0.1 cm}  ,           
\end{equation}
thus supporting the picture based on the 
 dominance of small parton separations. 
But 
a probe that, unlike $F_2$,  couples directly to gluons 
will give a  much better way of exploring 
this picture. 
The vector  meson cross sections considered in this paper,  
Eqs.~(\ref{photofact}) and 
(\ref{leptofact}),   
 will allow one 
 to study precisely  the relation 
between the large size of the diffractive 
gluon distribution  in a proton and the dominance of small 
 transparency lengths.

Note that   
this     study 
may   also be critical for  deep inelastic scattering on nuclei, 
if information extracted from  diffractive processes is 
to be used to 
compare  approaches to nuclear shadowing~\cite{nuclshad}.

It is  worth commenting   on the  $x_\pom$ dependence. 
In conventional fits to diffractive  data, 
a simple factorizing form of the $x_\pom$ and $\beta$ 
dependence  is often assumed.  
From the QCD point of view, 
this is an ansatz, not a  theory result.  
 In particular, notice  that  the diffractive gluon and quark 
distributions at low $\mu$ need not have the same 
  $x_\pom$ dependence. 
  Therefore, 
  even if a simple factorizing 
   form $x_\pom^{1-2 \alpha}$ is taken,   
it would be interesting  to 
 analyze the data using  two different behaviors   
 for the gluon and quark distributions,     
  parameterized by 
$\alpha_g$ and $\alpha_q$. The 
heavy meson  cross section discussed in this paper 
would   provide   a good way of determining~$\alpha_g$.

\bigskip

\acknowledgments

I  gratefully acknowledge  the 
 hospitality and support of 
the Institute for Nuclear Theory at the University of 
Washington while part of this work was being done. 
I thank    J.~Collins, L.~Frankfurt, 
D.~Soper and M.~Strikman for useful discussions.  
This research 
  is funded in part by the US Department of Energy 
 grant DE-FG02-90ER40577.


\bigskip



\begin{thebibliography}{999}
\bibitem{zeusexcl}
    ZEUS Collaboration  
    (J.~Breitweg {\it et al.}), Eur.\ Phys.\ J.\ C{\bf 6}, 603 (1999).       
\bibitem{h1chmn}
    H1 Collaboration (C.~Adloff {\it et al.}),   
    Eur.\ Phys.\ J.\ C{\bf 10}, 373 (1999).                     
\bibitem{h1excl}
    H1 Collaboration (C.~Adloff {\it et al.}),
    Phys.\ Lett.\ B {\bf 483}, 23 (2000). 
\bibitem{rys} 
    M.G.~Ryskin, Z.\ Phys.\ C {\bf 57}, 89 (1993); 
    M.G.~Ryskin, 
    R.G.~Roberts, A.D.~Martin and E.M.~Levin,  
    Z.\ Phys.\ C {\bf 76}, 231 (1997).   
\bibitem{brodetal} 
    S.J.~Brodsky, L.~Frankfurt, J.F.~Gunion, 
    A.H.~Mueller and M.~Strikman, Phys.\ Rev.\ D {\bf 50}, 3134 (1994).   
\bibitem{fskoepf} 
    L.~Frankfurt, W.~Koepf and  M.~Strikman,
    Phys.\ Rev.\ D {\bf 57}, 512 (1998). 
\bibitem{cfsfact} 
    J.C.~Collins, L.~Frankfurt  and M.~Strikman, 
    Phys.\ Rev.\ D {\bf 56}, 2982 (1997). 
\bibitem{leip}
    D.~M{\" u}ller, D.~Robaschik, B.~Geyer, 
    F.-M.~Dittes and J.~Ho{\v r}ej{\v s}i,  Fortsch.\ Phys.\  
    {\bf 42}, 101 (1994). 
\bibitem{xdji}
    X.D.~Ji,  Phys.\ Rev.\  D {\bf 55}, 7114 (1997). 
\bibitem{radyu}
    A.~Radyushkin, Phys.\ Lett.\ B {\bf 385}, 333 (1996).  
\bibitem{bjrest}
    J.D.~Bjorken, AIP Conference Proceedings No.~6, Particles 
    and Fields subseries No.~2 (New York 1972);  hep-ph/9601363; 
    J.D.~Bjorken, J.~Kogut and D.E.~Soper, Phys.\ Rev.\ D {\bf 3}, 1382 
    (1971).   
\bibitem{niko} 
    N.N.~Nikolaev and B.G.~Zakharov, Z.\ Phys.\ C {\bf 49}, 607 (1991), 
    {\em ibid} C      {\bf 53}, 331 (1992);   
    A.H.~Mueller,     Nucl.\  Phys.\ {\bf B335}, 115 (1990);  
    L.~Frankfurt and M.~Strikman,  Phys.\ Rep.\ {\bf 160}, 235 (1988). 
\bibitem{muedis98}
    A.H.~Mueller, in {\it Proceedings of the 
    International Workshop on Deep Inelastic Scattering DIS98} 
    (Brussels, April  1998),   
    World Scientific, Singapore 1998, p.3.  
\bibitem{ffss99} 
    L.~Frankfurt and M.~Strikman,  
    in {\it Proceedings of the 
    International Workshop on Deep Inelastic Scattering DIS99} 
    (Zeuthen), 
    Nucl.\ Phys.\ {\bf B} Proc.\ Suppl.\ {\bf 79},  671 (1999).         
\bibitem{hksnpb}
    F.~Hautmann, Z.~Kunszt and D.E.~Soper, 
    Nucl.\  Phys.\ {\bf B563}, 153 (1999).  
\bibitem{bere}
    A.~Berera and D.E.~Soper, Phys.\ Rev.\ D {\bf 53}, 6162 (1996). 
\bibitem{libby}
    S.B.~Libby and G.~Sterman,  Phys.\ Rev.\ D  {\bf 18}, 3252 (1978); 
    {\em ibid.} {\bf 18}, 4737 (1978). 
\bibitem{csrev} 
    J.C.~Collins and D.E.~Soper, 
    Ann.\ Rev.\ Nucl.\ Part.\ Sci.\ {\bf 37}, 383 (1987).  
\bibitem{diffcontour} 
    J.C.~Collins, Phys.\ Rev.\  D {\bf 57}, 3051 (1998), 
    (E) {\em ibid.} D {\bf 61}, 019902 (2000). 
\bibitem{psiphoto}
    E.L.~Berger and D.~Jones, Phys.\ 
    Rev.\ D {\bf 23}, 1521 (1981); R.~Baier and R.~R{\" u}ckl, 
     Z.\ Phys.\ C {\bf 19}, 251 (1983).  
\bibitem{niko94} 
    N.N.~Nikolaev and B.G.~Zakharov, Z.\ Phys.\ C {\bf 64}, 631 (1994).  
\bibitem{barniko}
    J.~Bartels, H.~Jung and M.~W{\" u}sthoff, 
    Eur.\ Phys.\ J.\ C{\bf 11}, 111 (1999). 
\bibitem{zerw} 
    M.~Kr{\" a}mer,  J.~Zunft, J.~Steegborn and 
     P.M.~Zerwas, 
         Phys.\ Lett.\ B {\bf 348}, 657 (1995); 
     M.~Kr{\" a}mer,  Nucl.\ Phys.\ {\bf B459}, 3 (1996). 
\bibitem{caccia}
       M.~Cacciari and M.~Kr{\"a}mer, 
       Phys.\ Rev.\ Lett.\  {\bf 76}, 4128 (1996); 
     P.~Ko, J.~Lee and H.S.~Song, 
     Phys.\ Rev.\ D {\bf 54}, 4312 (1996), 
     (E) {\em ibid.} D {\bf 60}, 119902 (1999); 
    J.~Amundson, S.~Fleming and I.~Maksymyk,  
    Phys.\ Rev.\ D {\bf 56}, 5844 (1997). 
\bibitem{albrow} 
    M.G.~Albrow,  in {\it Proceedings of the 
    International Symposium on Multiparticle Dynamics}, Tihany, Hungary, 
    October 2000, hep-ph/0102092. 
\bibitem{hscoltrans}
    F.~Hautmann and D.E.~Soper, Phys.\ Rev.\  D {\bf 63}, 011501 
    (2000). 
\bibitem{hkslett}
    F.~Hautmann, Z.~Kunszt and D.E.~Soper, 
    Phys.\ Rev.\ Lett.\      {\bf 81}, 3333 (1998). 
\bibitem{buchmue}
    W.~Buchm{\" u}ller,  T.~Gehrmann and A.~Hebecker, 
    Nucl.\ Phys.\ {\bf B537}, 477 (1999). 
\bibitem{golec}
    K.~Golec-Biernat and M.~W{\" u}sthoff, 
    Eur.\ Phys.\ J.\ C{\bf 20}, 313 (2001).  
\bibitem{zeusf2}
    ZEUS Collaboration  
    (J.~Breitweg {\it et al.}), Eur.\ Phys.\ J.\ C{\bf 6}, 43 (1999);  
    H1 Collaboration (C.~Adloff {\it et al.}),   
    Z.\ Phys.\  C{\bf 76}, 613 (1997).    
\bibitem{nuclshad}
    L.~Frankfurt, V.~Guzey, M.~McDermott and M.~Strikman, 
    hep-ph/0201230. 
\end{thebibliography}
\end{document}